\documentclass[]{article}
\usepackage{multicol}
\usepackage{apj}

\input epsf.sty

\newcommand{\mincir}{\raise -2.truept\hbox{\rlap{\hbox{$\sim$}}\raise5.truept
\hbox{$<$}\ }}
\newcommand{\magcir}{\raise -2.truept\hbox{\rlap{\hbox{$\sim$}}\raise5.truept
\hbox{$>$}\ }}
\newcommand{\siml}{\raise -2.truept\hbox{\rlap{\hbox{$\sim$}}\raise5.truept
\hbox{$<$}\ }}
\newcommand{\simg}{\raise -2.truept\hbox{\rlap{\hbox{$\sim$}}\raise5.truept
\hbox{$>$}\ }}

\newcommand{\be}{\begin{equation}}
\newcommand{\ee}{\end{equation}}
\newcommand{\ba}{\begin{eqnarray}}
\newcommand{\ea}{\end{eqnarray}}
\newcommand{\brr}{\begin{array}}
\newcommand{\err}{\end{array}}
\newcommand{\bc}{\begin{center}}
\newcommand{\ec}{\end{center}}

\begin{document}


\vspace{15mm}
\begin{center}
\uppercase{On the cosmological evolution of BL Lacs}\\
\vspace*{1.5ex} {\sc A. Caccianiga$^{1,2}$, T. Maccacaro$^2$, A. Wolter$^2$,
R. Della Ceca$^2$, I. M. Gioia$^{3,4}$}\\
\vspace*{1.ex}
{\small
$^1$ Observat\'orio Astron\'omico de Lisboa, Tapada da Ajuda, 1349-018, 
Lisboa, Portugal\\
E-mail: caccia@oal.ul.pt\\
$^2$ Osservatorio Astronomico di Brera, via Brera 28, 20121 Milano, Italy\\
$^3$ Institute for Astronomy, 2680 Woodlawn Drive, Honolulu, 
HI, 96822 USA\\
$^4$ Istituto di Radio Astronomia del CNR, via Gobetti 101, 
40129 Bologna, Italy \\
}
\end{center}
\vspace*{-6pt}
\begin{abstract}
We study the cosmological evolution of BL Lac objects by applying 
the $V_e/V_a$ analysis to a new sample of 55 objects presented 
for the first time in this paper. This sample has been selected 
from the 239 sources with the brightest X-ray flux ($\geq$4$\times$10$^{-13}$
erg s$^{-1}$ cm$^{-2}$)
and relatively bright optical counterpart (B$\leq$20.5)
among the $\sim$1600 objects included in the REX survey. 
The uniform distribution
of the $V_e/V_a$ values found in the sample suggests that BL Lac objects
are not affected by a strong cosmological evolution in contrast with
the behavior observed in the emission line AGNs. The $V_e/V_a$ analysis
applied to the subsample of the High energy peaked BL Lacs (HBL) does not
reveal a significant departure from a uniform distribution.  
This result suggests either that the cosmological evolution is
less extreme than that previously found in other samples or
that it is carried out only by a minority of objects, namely by 
the most extreme HBLs.

\vspace*{6pt}
\noindent

{\em Subject headings:} 
surveys - galaxies: active - quasar: general - BL Lacertae 
objects: general
\end{abstract}

\begin{multicols}{2}

\section{Introduction}

Among the Active Galactic Nuclei (AGN), BL Lac objects represent
a minority characterized by an almost  featureless optical 
spectrum and by an intense ``blazar'' activity (variability, flat 
radio spectrum, polarization).  
Due to their scarcity, the study of BL Lacs is 
not simple and many important issues, like the cosmological evolution, 
are not yet well understood.  
The first attempts to estimate the evolutionary behavior of BL Lac objects
were based on the {\it Einstein} Medium Sensitivity Survey (MSS, Maccacaro
et al. 1982) and its extension (EMSS, Gioia et al. 1990).  
Evidence for negative evolution (objects
less numerous/luminous in the past than now) was found 
(Maccacaro et al. 1984; Wolter et al. 1991; Morris et al. 1991; 
Wolter et al. 1994; Rector et al. 2000). At the same time,
the analysis of the 1 Jy radio selected sample of BL Lacs found
a mildly positive evolution (Stickel et al. 1991; Rector \& Stocke 2001) 
at odds with the X-ray results. However, all these conclusions 
were based on small samples containing 20--40 objects at most.

The recent availability of large radio and X-ray surveys boosted
the attempts at selecting large samples of BL Lacs. Different 
techniques are used in order to increase the efficiency of the 
selection process: a very efficient technique turned out to be the 
combination of radio and X-ray information thanks to the fact that 
BL Lacs are both X-ray and radio loud sources as firstly pointed 
out by Stocke et al. (1991). This technique reaches efficiencies 
of 15\%--30\% (e.g. Wolter et al. 1997)  
depending on the flux limits involved, while in a purely X-ray or 
radio survey the percentage of BL Lacs hardly reaches 5\% (e.g. 
Stocke et al. 1991). 
A  radio survey of flat spectrum sources, like the 1~Jy survey,
has an efficiency of about 12\% (Stickel et al. 1991) 
although deeper flat spectrum
radio surveys include higher percentages of BL Lacs
(up to 20\% in the 200~mJy sample, March\~a et al. 1996 and in the 
CLASS Blazar Survey, March\~a et al. 2001, Caccianiga et al. 2001). 

Thus, surveys based on
the cross-correlation of radio and X-ray catalogs have been initiated
(the DXRBS survey, Perlman et al. 1998, the RGB survey, Laurent-Muehleisen 
et al. 1998, the REX survey, Maccacaro et al. 1998; 
Caccianiga et al. 1999, hereafter Paper~I). 
So far, the general picture is still quite unclear. Some surveys
(Bade et al. 1998; Giommi, Menna \& Padovani 1999; Giommi et al. 2001) find  
negative evolution but only for a restricted sub-set of the BL Lac objects, 
namely the ones with a synchrotron emission peaked at
high energies (the High energy peaked BL Lacs, HBL). A continuous 
trend, ranging from slightly positive evolution for the
Low energy peaked BL Lacs (LBL) to strong negative evolution
for HBL has been found also by Rector \& Stocke (2001) after
the re-analysis of the 1~Jy and EMSS samples.
At the same time, preliminary results from other surveys, like the DXRBS 
and the REX survey, are not finding any 
cosmological evolution, neither negative nor positive 
(Padovani 2001; Caccianiga et al. 2001). 

In this paper we present a  study of the cosmological
evolution of BL Lacs based on an almost (91\%) complete sample 
of 55 BL Lacs created by imposing a relatively bright 
X-ray and optical cut on the REX survey. So far, this is the 
largest sample
with a spectroscopic classification that is used to estimate
the cosmological evolution of BL Lacs. 
The sample is presented in Section~2 while the statistical 
analysis is described in Section~3. In Section~4 we compare
the result with those reported in the literature. 
Finally, the discussion and the conclusion are presented
in Section~5.

\section{The X-ray bright REXs sample}

The REX survey is the 
result of a positional cross-correlation between the  
NRAO VLA Sky Survey (NVSS, Condon et al. 1998) at 1.4 GHz and
an X-ray catalogue of about 17,000 serendipitous sources detected
in 1202 pointed ROSAT PSPC fields (see Paper~I for details).
The good accuracy of the NVSS (VLA) positions is used to
pinpoint the optical counterpart. 
The spectroscopic identification of the sample is in progress.
So far, about 40\% of the sample has been spectroscopically 
identified and some
of the newly discovered objects have been already presented  
(Wolter, Ruscica \& Caccianiga 1998; Caccianiga et al. 2000).
The identification strategy that we have always adopted is to observe
X-ray bright sources first, so as to obtain a complete subsample 
(of increasing size) with a very high fraction of identified sources
usable for statistical studies. 
At present, at the X-ray flux limit of 4$\times$10$^{-13}$ 
erg s$^{-1}$ cm$^{-2}$ and considering only the objects with a 
blue magnitude brighter than 20.5$^{mag}$ 
the sample is almost completely identified (identification
rate of 91\%). We have thus analyzed this complete sample, 
that we call X-ray Bright REX (XB-REX), with the main goal
of studying the statistical properties of the BL Lac objects.
The incidence of the relatively bright X-ray cut-off on the
final results is discussed in Section~4.
Despite the fact that the REX survey has been created by using pointed
PSPC images characterized by different exposure times and, thus, 
different flux limits, at the bright X-ray flux limit considered
for the XB-REX sample the sky-coverage function is flat, i.e. 
the almost totality (99\%) of the surveyed area reaches the 
X-ray flux limit of 
4$\times$10$^{-13}$ erg s$^{-1}$ cm$^{-2}$. 
This greatly
simplifies the analysis of the sample which is characterized
by a single X-ray, radio and optical flux limit.

In total, the XB-REX sample contains 239 objects.
The magnitudes are preferentially taken from the APM (Automatic Plate
Measuring\footnote{http://www.ast.cam.ac.uk/$\sim$apmcat/}) 
database (O magnitude) or from the LEDA catalogue (Paturel et al. 1996).
For about 10\% of objects, for which no blue magnitude was available,
we have estimated it from the APM E magnitude\footnote{
We have assumed O -- E =  1.4 which is the mean value found
in the sample} or used the B 
magnitude found in the literature. 
In case of the BL Lac objects presented in this paper, the great majority
of the magnitudes are derived from APM except for 4  objects where the 
magnitudes are taken from the literature.

In the XB-REX sample 91\% of the objects (218 out of 239) are 
already identified, either from the literature or from our own 
spectroscopy. In total, 55 BL Lac objects (25\% of the identified objects), 
95 emission line AGNs (EL AGN) and 66 objects classified as ``galaxies'' 
are found. The complete sample of 55 BL Lacs is reported in 
Table~2. 
In addition, two objects that are classified as stars are
probably spurious optical counterparts and should be rejected from
the sample (the real counterpart is likely to be fainter than
the optical limit used for the XB-REX sample).

The classification of the different classes follows the
rules described in Paper~I.
In case of BL Lac objects, we have used the ``expanded'' 
definition based on the Ca II contrast at 4000\AA\ ($\Delta$)
which must be below 40\% instead of 25\% as firstly proposed
by  Stocke et al. (1989). This
modification is justified by the evidence 
that the limit of 25\% is too restrictive and can miss some true
BL Lac objects (March\~a et al. 1996, Laurent-Muehleisen et al. 1998, 
Rector et al. 2000).
In principle, this criterion may 
introduce in the sample some ``optically passive'' galaxies 
for which $\Delta$ is low due, for instance, to a
particular evolutionary stage of the galaxy rather than to
the presence of a BL Lac nucleus. Nevertheless, we have estimated 
(see Paper~I) that the percentage of ``normal'' galaxies
in the sample of BL Lacs defined according to the above definitions
should be less than 10\%, which means less than 5 spurious objects in
the sample. A VLA survey of the BL Lacs and ``optically passive''
galaxies discovered in the REX survey is in progress and should
provide a compelling  tool to separate BL Lac nuclei from 
``normal'' galaxies. 
The redshift distribution of the 37 BL Lac objects for which a
redshift has been measured is presented in Figure~1.

By assuming that the percentage of BL Lacs among the
unidentified objects is the same as for the identified 
sources (i.e. 25\%) we expect that only 5 BL Lacs will
be found after the completion of the identification
process. This number is in complete agreement with the 
one computed by
comparing the overall spectral energy distribution
of the  unidentified objects with that of  the 55 BL Lacs.
Through this comparison we predict only  3--4 missing  BL Lacs.
Given the small number of the expected
missing objects, the conclusions of the analysis presented in
this paper are not likely to  change after the completion of
the identification process.

\includegraphics{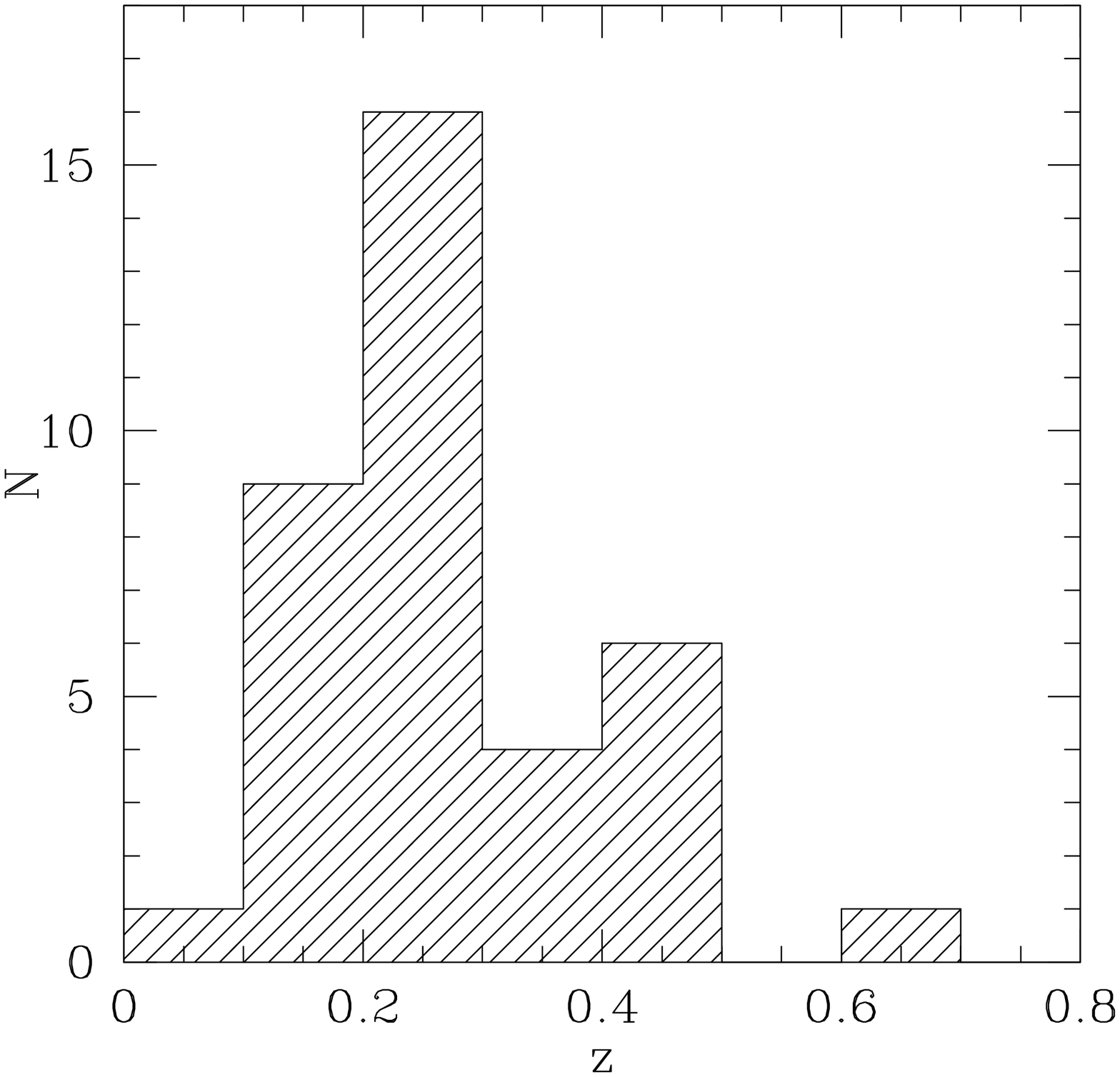}
$\ \ \ \ \ \ $\\
\vspace{5.8truecm}
$\ \ \ $\\
{\small\parindent=3.5mm {Fig.}~1.---
The redshift distribution of the 37 BL Lac objects in the
XB-REX sample with a spectroscopic measured redshift.}\vspace{5mm}

\section{The Cosmological evolution: Results}

The high identification level  of the XB-REX sample allows us to apply the 
$V_e/V_a$ test  extended to the multifrequency 
selected samples (Avni \& Bachall 1980). 
This test checks the uniformity of the spatial distribution
of the objects in a flux-limited sample. A departure from a
uniform distribution is interpreted as due to an evolutionary
behavior of the population under study. For each object,
the two volumes $V_e$ and $V_a$ are derived on the basis of 
the observed redshift (z) and the maximum redshift (z$_{max}$) at which the 
object would produce a flux equal to the flux limit of the survey.
Since the XB-REX is characterized by 3 flux limits, the most
stringent value of z$_{max}$ is considered, i.e.:

\begin{center}
z$_{max}$ = min (z$_{max}^R$, z$_{max}^O$, z$_{max}^X$)
\end{center}

\noindent where z$_{max}^R$, z$_{max}^O$, z$_{max}^X$ are the maximum redshifts
obtained respectively in the radio, optical and X-ray band. 
If the $V_e/V_a$ values 
are not uniformly distributed between 0 and 1,  there are indications
for either an evolutionary behavior of the population under study
or for the presence  of selection effects biasing the sample. 
In particular the first requirement for a uniform distribution
is that the average of the $V_e/V_a$ values is consistent with
0.5. The effective uniformity of the distribution must then be checked
with, for instance, a Kolmogorov-Smirnov test. 
By using the sample of 55 BL Lac, we have computed the $V_e/V_a$ values
for each source. 
For the objects without a redshift, 
we have assumed z=0.27 (which is the mean redshift of the sample). 
We have also assumed 
$\alpha_R$ = 0 and $\alpha_X$ = 1 for the spectral indices 
in the radio and in the X-ray band, respectively. In the optical
band, BL Lacs show a wide range of spectral indices, with  
radio selected BL Lacs being steeper than X-ray selected BL Lacs 
(Falomo, Scarpa \& Bersanelli 1994). Since the majority of our
sources are more similar to the X-ray selected objects, 
we have used the mean value found by Falomo, Scarpa \& Bersanelli (1994) 
for these sources ($\alpha_O$ = 0.65).  

The resulting value of the $<V_e/V_a>$, reported in 
Tab.~1 ($<V_e/V_a>$=0.484$\pm$0.039), 
is fully consistent with 0.5.
The K-S test shows that the $V_e/V_a$ values are consistent with a
uniform distribution between 0 and 1 (probability=76.3\%). 

In order to check the stability of the results, we have performed  
the analysis under different assumptions. 
If we assume, for all the objects without a redshift determination, 
the values taken from the extremes of the observed redshift distribution
(Fig~1), i.e. 0.1 and 0.7, we obtain $<V_e/V_a>$ = 0.47 and 0.50 
($\pm$0.039) respectively.  
Thus, the result is not significantly affected by the 
number of objects without a redshift determination. 

The  $V_e/V_a$ analysis may also be affected by a wrong assumption 
on the spectral slope. The expression used for the 
K-correction ($K=(1+z)^{(\alpha-1)}$), based on the assumption of 
a power-law spectrum with spectral index 
$\alpha$, can mimic a luminosity evolution (or cancel a real one)
if the value of $\alpha$ is over or underestimated.
By assuming extreme values for the $\alpha$
in the 3 bands, namely 
$\alpha_X$=0.5, $\alpha_R$=--0.5, $\alpha_O$=0.0 and
$\alpha_X$=2, $\alpha_R$=1, $\alpha_O$=2, 
the $<V_e/V_a>$ values range from 0.47 to 0.52.
We conclude that the K-correction does not play a fundamental role
in the analysis unless very extreme (and unrealistic) values of
$\alpha$ are used.
\vspace{6mm}
\hspace{-4mm}
\begin{minipage}{9cm}
\renewcommand{\arraystretch}{1.5}
\renewcommand{\tabcolsep}{1mm}
\begin{center}
\vspace{-3mm}
~\\ ~\\
TABLE 1\\
\vspace{2mm}
{\sc Results of the $V_e/V_a$ analysis\\}
\footnotesize
\vspace{2mm}
\begin{tabular}{l l l l r}
\hline\hline \\
Sample (N. objects)& $<V_e/V_a>$ & Error$^1$ & s$^2$ & 
KS prob.(\%)$^3$ \\
Total (55)                   & 0.484 & 0.039 & 0.4  & 76.3 \\
``classical'' BL (46)        & 0.461 & 0.043 & 0.9  & 42.6 \\      
$\alpha_{OX}\leq$0.91 (22)   & 0.488 & 0.061 & 0.2  & 82.2 \\
$\alpha_{RX}\leq$0.62 (25)   & 0.469 & 0.058 & 0.5  & 56.2 \\
EL AGNs (95)                 & 0.630 & 0.030 & 4.3  & $<<$0.1 \\
simulated sample (64)        & 0.423 & 0.036 & 2.1  &  9.7 \\
\hline
\end{tabular}\\
\end{center}
\footnotesize{}

$^1$ Error = 1/$\sqrt(12N)$;

$^2$ The significativity defined as the distance from 0.5 in units of 
$\sigma$;

$^3$ The KS probability (null hypothesis = the $V_e/V_a$ distribution is
uniform bewteen 0 and 1).
\end{minipage}
\vspace{3mm}

We have then investigated whether different types of objects 
show different cosmological behaviors. 

{\bf ``Classical'' BL Lacs}. 
We have computed the $<V_e/V_a>$ by using only the ``classical''
BL Lacs, i.e. defined with the most restrictive limit on  
$\Delta$ (25\%). The 46 ``classical'' BL Lacs have a $<V_e/V_a>$
lower than the entire sample but still consistent with 0.5 
($<V_e/V_a>$=0.461$\pm$0.043).

{\bf HBLs}
We have considered  only the 22 BL Lacs 
with an ``extreme'' X-ray/optical ratio ($\alpha_{OX}\leq$0.91)
typical of HBLs. 
We have also divided the sample according to the ratio between the
X-ray and the radio flux and considered only those with 
$\alpha_{RX}\leq0.62$.
In both cases the $<V_e/V_a>$ value is consistent with 0.5 
($<V_e/V_a>$=0.488$\pm$0.061 and 0.469$\pm$0.058 respectively). 

{\bf Emission Line AGNs}. 
For comparison, we have computed the $V_e/V_a$ for the Emission Line
AGNs found in the XB-REX sample. 
By using the sample of 95 EL AGN, we have  computed the $<V_e/V_a>$  
with the following assumptions: $\alpha_R$ = 0.4 
(the mean value found for the emission line AGNs in the REX survey, 
Caccianiga et al. 2000), $\alpha_X$ = 1 and $\alpha_O$ = 0.65. 
The result is indicative of a strong positive evolution  
($<V_e/V_a>$=0.630$\pm$0.030). 
We have then estimated the evolution by assuming a 
pure luminosity evolution of the form: $L(z) = L(0)(1+z)^k$
under the hypothesis that the values of $k$ in the three bands 
(radio, optical and X-ray) are the same. 
The best fit value is $k=$3.0 with
a 1$\sigma$ interval of (2.5, 3.3). 
The value of $k$ is intermediate between that found 
for the X-ray selected  ($k$=2.92, Della Ceca et al. 1994)
and the optically selected (e.g. $k$=3.49, Padovani 1993) 
radio-loud AGNs. Both values are consistent within 1$\sigma$ with 
the one found in the XB-REX sample.

\section{The simulations}

The absence of cosmological evolution in BL Lacs
is potentially at odds with the results 
presented in the literature, claiming the existence of 
negative evolution (Morris et al. 1991; Wolter et al. 1994;
Bade et al. 1998; Giommi, Menna \& Padovani 1999; Rector et al. 2000)
or a positive evolution (Stickel et al. 1991; Rector \& Stocke 2001). 
Assuming a luminosity evolution of the form:
\begin{equation}
L(z) = L(0) exp (c \tau)
\end{equation}
where $\tau$ is the look-back time, our result is consistent, 
at 1$\sigma$ confidence level, with a value of $c$ between +2.0 and --3.5.
Thus, both the negative ($c$=--7.0, Wolter et al. 1994) and positive 
($c$=3.1, Stickel et al. 1991) best-fit parameters found in the
literature fall outside the 1$\sigma$ range.  

It can be argued that the negative evolution of BL Lacs is
not visible in the XB-REX sample simply because this sample is not 
deep enough. The flux limit of 4$\times$10$^{-13}$ 
erg s$^{-1}$ cm$^{-2}$ in the 0.5-2.0 keV band 
corresponds to a flux limit of 7.2$\times$10$^{-13}$ erg s$^{-1}$ cm$^{-2}$ 
in the 0.3--3.5 keV band (which is the EMSS energy band) 
assuming $\alpha_x$=1. 
This flux is higher than the lowest limit reached by
the EMSS complete sample (5$\times$10$^{-13}$ erg s$^{-1}$ cm$^{-2}$)
used by Morris et al. (1991)\footnote{the flux limit 
of  5$\times$10$^{-13}$ erg s$^{-1}$ cm$^{-2}$ is actually reached 
only by a fraction (58\%) of the area covered by the EMSS survey. 
The flux limit corresponding to the 90\% of the area is much
higher ($\sim$2.5$\times$10$^{-12}$ erg s$^{-1}$ cm$^{-2}$)}.
The flux limit of the XB-REX is thus in the 
region where the BL Lac LogN-LogS measured in EMSS survey 
(Wolter et al. 1991) starts to flatten. Thus, the negative  
evolution found in the EMSS sample might simply not be observable in 
the XB-REX sample. 

In order to verify this hypothesis we have analyzed a simulated sample
(see Paper~I) created under the assumption of
negative evolution of the exponential form given by Eq.~1
with the best-fit value of  $c=-7.0$ found by Wolter et al. (1994).
If we apply the $V_e/V_a$ analysis to the entire simulated sample and
we compute the evolutionary parameter, we find a best-fit
value of $-$6.8, i.e. in excellent agreement with the input
value. This first test demonstrates the validity of the method 
when applied to a multifrequency survey. 

If we apply to the simulated sample the same radio, optical and 
X-ray flux limits used for the XB-REX, 
the resulting sample contains 65 objects. This number is in excellent
agreement with the observed number of objects if we consider that
10\% of the objects are expected to be missed in the positional
radio/X-ray correlation as explained in Paper~I and
an additional 9\% is expected to be missed due to the identification level
(91\%) achieved so far in the XB-REX sample. Considering these 
corrections, the number of BL Lacs expected in the XB-REX sample
is about 0.9$\times$0.91$\times$65 = 53 that should  be compared with 
the observed number of 55.

The $<V_e/V_a>$ analysis of the simulated sample shows the presence of 
the negative evolution ($<V_e/V_a>$= 0.423 $\pm$ 0.036, see Tab.~1). 
The $V_e/V_a$ distribution is not uniform although 
the KS test rejects the hypothesis of no-evolution only 
marginally (probability of 9.7\%). By assuming
the same form of the evolution used as input for the
simulations, we obtain a best-fit value ($c$=--4.5 with 1$\sigma$
range of --7.2, --2.2) which is lower than the input value (--7.0)
although consistent at 1$\sigma$. 
Thus, the negative evolution is recovered
but it is weaker than the one given as input. This result
is probably due to the fact that, as the flux limit increases,
the negative evolution becomes less and less evident
since the euclidean part of the LogN-LogS is 
sampled. 

In conclusion our simulations show that, at the X-ray flux limit of 
the XB-REX sample, the negative evolution seen in the EMSS sample should be
less extreme but still detectable.
The result of the analysis presented here may indicate that, 
if present, the negative evolution is  weaker than that estimated from
the EMSS sample.

\section{Discussion and conclusions}

We have presented the result of the $V_e/V_a$ test 
on a  sample of 55 BL Lacs extracted from the REX survey. 
No evidence of positive or negative evolution is found. 
The analysis presented in this paper is rather stable, i.e.
it is not affected much by the fact that some sources do not have
a redshift determination  (indeed, the $V_e/V_a$ 
test does not depend very much on the redshift of the sources). 

Simulations based on the EMSS sample has demonstrated
that, at the X-ray flux limit of the XB-REX sample, the 
negative evolution found by Wolter et al. (1994) is expected
to be less evident although still detectable.
More puzzling is the discrepancy between the results found 
in the XB-REX sample and the strong negative evolution presented 
in Bade et al. (1998). Since the RASS sample has a higher  X-ray flux
limit than the XB-REX sample, the strong negative evolution found in 
the RASS sample  should have been clearly detected in
our analysis. 

A possible explanation of this apparent contradiction is that 
the evolutionary behavior is not the same for all the
different sub-classes of BL Lacs. As suggested by different
authors, there may be a continuous trend of the evolutionary
properties within the BL Lac class, with the most extreme HBLs
evolving negatively, from one side, and the most extreme LBLs 
evolving positively, from the other side. Depending on the 
flux limit(s) of a given survey and/or to its selection criteria,
a sample may contain a different percentage of HBLs and LBLs and
this can produce different results in the $V_e/V_a$ analysis. In Figure~2 
the $\alpha_{RX}$ distribution of the 55 BL Lacs selected in the
XB-REX sample (tick line) is presented and compared to the same distribution
for the EMSS (shaded histogram on the left side), the 1~Jy (filled histogram 
on the right side) and the RASS (dashed line) BL Lac samples. 
It is clear that the 
XB-REX BL Lacs are covering a range of $\alpha_{RX}$ which is similar
to that observed in the X-ray selected samples (EMSS and RASS) although
it does not reach the  very low end of the distribution ($\alpha_{RX}<$0.5). 
Hence,  we may expect qualitatively that the 
evolutionary behavior of the XB-REX BL Lacs is  similar to
that observed in the X-ray selected samples but less extreme since the
very low $\alpha_{RX}$ objects are not included in the sample. 
Since we do not observe evidence for evolution even when we consider the
objects with $\alpha_{RX}\leq$0.62, the conclusion is that the extreme
negative evolution must be carried out mainly by the minority of the 
objects with $\alpha_{RX}$ most extreme, not selected in our sample.
Surveys more specifically  
designed to find the extreme HBLs, like the Sedentary multifrequency 
survey, should be in the best position to observe the signs of a 
negative evolution (Giommi et al. 2001). 

\includegraphics{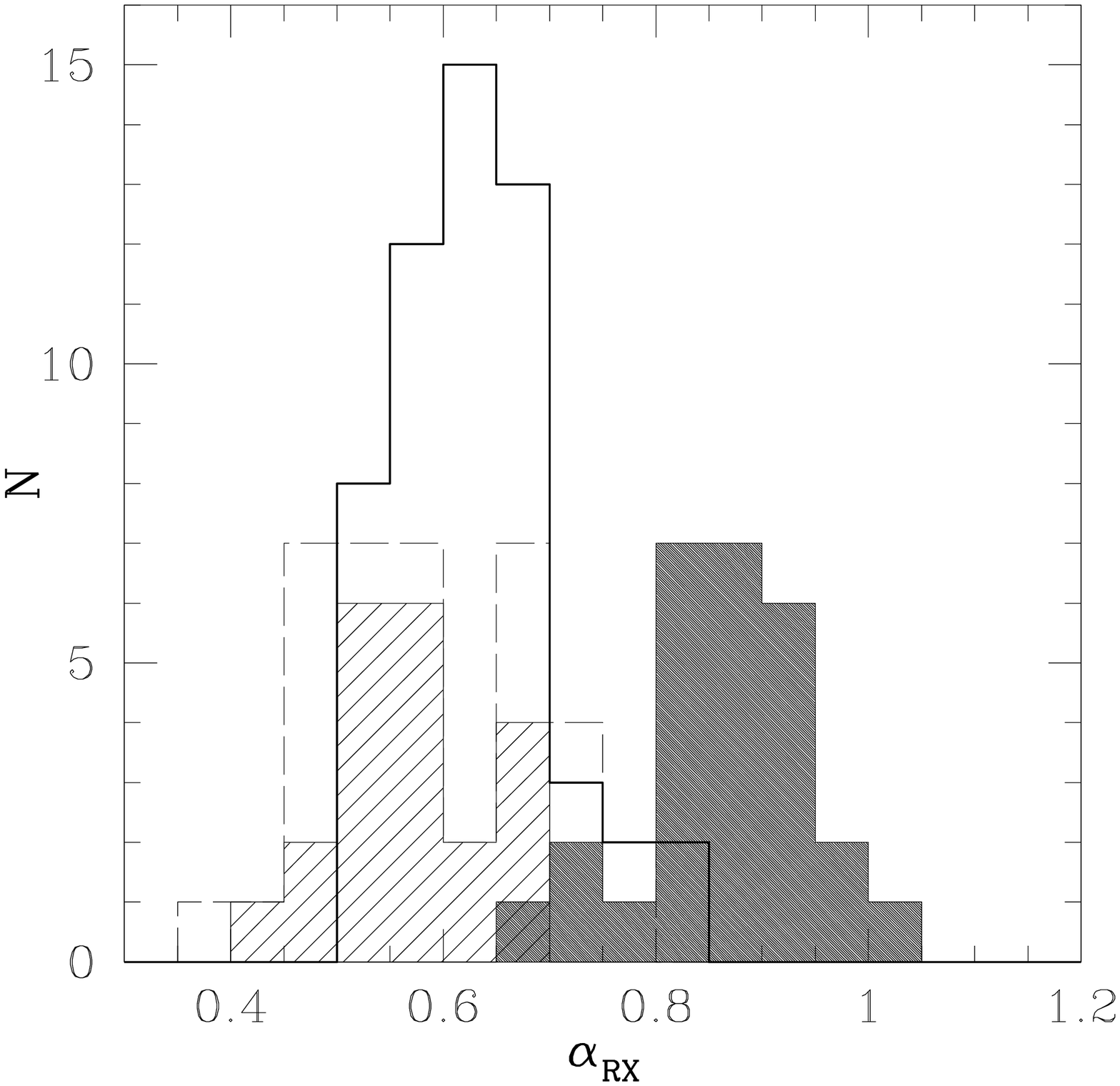}
$\ \ \ \ \ \ $\\
\vspace{5.8truecm}
$\ \ \ $\\
{\small\parindent=3.5mm {Fig.}~2.---
The radio-to-X-ray spectral index ($\alpha_{RX}$) distribution of the 
BL Lac objects in different samples: XB-REX (tick line), EMSS 
(shaded histogram on the left side), 1Jy (filled histogram on the 
right side) and RASS (dashed line)}\vspace{5mm}

In any case, the analysis presented in this paper confirms that BL Lac 
objects do not follow the same evolutionary behavior of the emission line
AGNs. Possible explanations for this difference 
have been recently proposed  by Cavaliere \& Malquori (1999)
on the basis of a 
different accretion rate on the central massive black-hole:
BL Lac objects may be characterized by a low accretion rate
($\dot m = \dot M\,c^2/L_E \sim$10$^{-2}$) that produces a 
very mild luminosity evolution
in contrast with what happens in the Flat Spectrum Radio Quasars (FSRQ), 
where the
accretion is close to the Eddington limit ($\dot m \sim $1) producing
a much faster luminosity evolution. The predicted time scale
for the luminosity evolution in the BL Lac population is expected 
to give a $<V_e/V_a>$ consistent with 0.5, as observed in the 
XB-REX sample.
Superimposed to this
effect, a possible transition from a FSRQ into a BL Lac object,
due to a reduction of the accretion rate at low redshifts,
may cause negative (density) evolution in
BL Lac objects (objects more numerous now than at high redshift)
as described in Cavaliere \& D'Elia (2001). 
A similar evolutionary link between FSRQ and BL Lac has been 
proposed by B\"ottcher \& Dermer (2001) where the reduction 
of the black-hole accretion power with time may switch a FSRQ
into a BL Lac object following the sequence 
FSRQ$\rightarrow$LBL$\rightarrow$HBL. 
As discussed before, the analysis of the XB-REX
sample indicates that only a mild negative evolution
is consistent with the data although only a narrow range of 
BL Lac ``flavors'' is actually covered in this sample (see Fig.~2). 
After the completion of identification process of the entire 
REX survey, which contains 7 times more sources than the XB-REX 
sub-sample, a wider range of BL Lac types will be covered allowing
a more complete study of the evolutionary behaviors within the
BL Lac class.

\acknowledgments  
We thank M.J. March\~a, P. Giommi, J. Stocke and an anonymous 
referee for enlightening discussions and suggestions. 
This research has made use of the NASA/IPAC extragalactic database (NED), 
which is operated by the JET Propulsion Laboratory, Caltech under contract 
with the National Aeronautics and Space Administration.
This work has received partial financial support from the Portuguese 
FCT (PRO15132/1999), from the Italian Space Agency (ASI) and
MURST (COFIN00-02-36).

\end{multicols}

\vspace{6mm}
\hspace{-4mm}
\begin{minipage}{20cm}
\renewcommand{\arraystretch}{1.5}
\renewcommand{\tabcolsep}{1mm}
\begin{center}
\vspace{-3mm}
~\\ ~\\
TABLE 1\\
\vspace{2mm}
{\sc The XB-REX sample of BL Lacs\\}
\footnotesize
\vspace{2mm}
\begin{tabular}{l r r l l r l l l}
\hline\hline \\
Name & f$_X$ & S$_{1.4~GHz}$ & blue mag & z & obs  & 
other name & type & $V_e/V_a$\\

\hline

 1REXJ002727+2607.1  &  7.76 & 56.5 & 19.0$^E$ &0.36$^t$&Y &                                  &     & 0.48\\
 1REXJ012430+3249.7  & 12.81 & 14.7 & 19.9     & --    & Y &                                  &     & 0.53\\
 1REXJ012656+3307.4  & 26.57 &  6.9 & 17.7     & --    & Y &                                  &     & 0.65\\
 1REXJ015307+7517.7  &  5.75 & 21.6 & 18.9$^E$ & --    & Y &                                  &     & 0.65\\
 1REXJ020106+0034.0  & 37.69 & 13.2 & 18.5     & 0.299 &   & MS0158.5+0019                    &     & 0.29\\
 1REXJ021905$-$1725.2& 15.53 & 62.5 & 17.2     & 0.128 & Y & XRS J0219-1725                   &BLC  & 0.18\\
 1REXJ022239+4302.1  & 29.08 &2303.6& 17.0     & 0.444 &   & HB0219+428                       &     & 0.15\\
 1REXJ031006+4056.8  &  6.41 &  30.0& 18.4$^E$ &0.137$^t$& Y&                                 &     & 0.54\\
 1REXJ033312$-$3619.7&  8.07 &  14.7& 19.5     & 0.308 &   & MS0331.3-3629                    &     & 0.45\\
 1REXJ033812$-$2443.8& 12.65 &  13.9& 19.6     & 0.2509 &   & E0336-248                        &     & 0.37\\
 1REXJ035305$-$3623.1& 17.02 &   6.6& 19.0     & --    & Y &                                  &     & 0.69\\
 1REXJ050639$-$0858.0&  5.75 &  21.0& 18.9     & --    & Y &                                  &     & 0.65\\
 1REXJ053629$-$3343.0& 24.75 &  94.7& 17.0     & --    & Y &                                  &     & 0.13\\
 1REXJ054656$-$2204.9&  7.23 &  12.3& 20.3     & 0.247 & Y &                                  &BLC  & 0.76\\
 1REXJ055806$-$3838.4& 79.47 & 105.7& 17.7$^V$ & --    &   & EXO0556.4-3838                   &     & 0.06\\
 1REXJ062444$-$3230.8&  6.03 &  44.0& 18.5     & 0.275 &   & WGAJ0624-3231                    &     & 0.62\\
 1REXJ073329+3515.7  & 15.84 & 102.3& 17.9     & 0.177 & Y &                                  & BLC & 0.19\\
 1REXJ074405+7433.9  & 72.55 &  23.3& 16.6     & 0.315 &   & MS0737.9+7441                    &     & 0.15\\
 1REXJ074722+0905.8  & 17.86 &  40.3& 19.0     & --    & Y & RGB J0747+090                    &     & 0.20\\
 1REXJ080102+6444.8  &  5.80 &  12.7& 19.7     & 0.20$^t$&Y& RXJ0801.1+6444                   &     & 0.63\\
 1REXJ080526+7534.4  & 17.25 &  52.6& 18.1$^B$ & 0.121 &   & RXJ08054+7534                    &     & 0.15\\
 1REXJ081054+4911.0  &  4.05 &  11.3& 17.0$^E$ & 0.115 & Y &                                  & BLC & 0.98\\
 1REXJ081421+0857.0  &  4.43 &  11.8& 19.9     & 0.23$^t$&Y&                                  & BLC & 0.88\\
 1REXJ081917$-$0756.4&  9.61 &   7.2& 18.5     & --    & Y &                                  &     & 0.62\\
 1REXJ082706+0841.3  &  6.03 &  15.3& 20.4$^E$ & --    & Y &                                  &     & 0.89\\
 1REXJ091552+2933.4  & 14.33 & 342.0& 16.3     & --    &   & HB0912+297                       &     & 0.23\\
 1REXJ093055+3503.6  &  5.40 & 484.4& 20.4     & --    & Y & 87GB082752.8+351642              &     & 0.91\\
 1REXJ094022+6148.4  & 29.98 &  12.8& 19.2     & 0.209$^1$ &Y& RX J0940.3+6148                & BLC & 0.29\\
 1REXJ101616+4108.1  & 22.84 &  14.8& 18.9     & 0.282$^2$& Y &  FBQS J101616.8+410812        & BLC & 0.25\\
 1REXJ103335$-$1436.4& 13.37 &  12.9& 20.2     & 0.367 & Y &                                  &     & 0.73\\
 1REXJ111158+4856.9  &  9.16 &  45.9& 20.3     & --    & Y &                                  &     & 0.78\\
 1REXJ120412+1145.9  & 27.43 &  15.1& 17.7     & 0.46$^t$&Y&  1RXS J120413.0+114549           &     & 0.27\\
 1REXJ121026+3929.1  & 15.56 &  19.0& 20.3     & 0.615 &   & HB1207+397                       &     & 0.80\\
 1REXJ121510+0732.0  & 25.25 & 137.7& 17.8     & 0.130 &   & RGBJ1215+075                     &     & 0.10\\
 1REXJ121752+3007.0  & 28.05 & 572.7& 15.4     & 0.237$^3$&& HB1215+303                       &     & 0.11\\
 1REXJ122121+3010.6  &219.97 &  71.5& 16.7     & 0.130$^4$&& PG1218+304                       &     & 0.03\\
 1REXJ123133+1421.3\footnote{
This object has been presented in Wolter et al. (1997) with the name
REXJ1231.4+1421.
}  
                     & 12.00 & 55.5 & 18.88    & 0.26  & Y &                                  &     & 0.28\\
 1REXJ124141+3440.4  & 10.22 & 10.2 & 19.6$^B$ & --    &   & FIRSTJ124141.3+34403             &     & 0.40\\
 1REXJ124232+7634.4  &  8.14 &  8.6 & 18.9     & --    & Y &                                  &     & 0.49\\
 1REXJ125134$-$2958.7& 14.54 & 10.4 & 19.1     & 0.487 &   & 1ES1248-296                      &     & 0.42\\
 1REXJ125359+6242.9  &  5.96 & 12.0 & 18.3     & --    & Y &                                  &     & 0.62\\
 1REXJ131155+0853.7  & 17.49 &  5.3 & 20.3     & 0.480$^{5}$ & Y &    RX J1311.9+0853         & BLC & 0.93\\

\hline
\end{tabular}\\
\end{center}
\footnotesize{}

\end{minipage}
\vspace{3mm}

\newpage

\vspace{6mm}
\hspace{-4mm}
\begin{minipage}{20cm}
\renewcommand{\arraystretch}{1.5}
\renewcommand{\tabcolsep}{1mm}
\begin{center}
\vspace{-3mm}
~\\ ~\\
TABLE 1\\
\vspace{2mm}
{\sc (continued)\\}
\footnotesize
\vspace{2mm}
\begin{tabular}{l r r l l r l l l}
\hline\hline \\
Name & f$_X$ & S$_{1.4~GHz}$ & blue mag (flag) & z & obs  & 
other name & type & $V_e/V_a$ \\

\hline
 1REXJ132532+6621.0  &  4.44 &  5.1 & 19.4     & 0.210 & Y &                                  & BLC & 0.97\\
 1REXJ133529$-$2950.6& 12.61 & 10.6 & 19.7$^V$ & 0.256 &   & MS1332.6-2935                    &     & 0.41\\
 1REXJ134105+3959.8  & 51.69 & 88.8 & 18.5$^{B}$&0.163&  & B31338+402                       &     & 0.10\\
 1REXJ140923+5939.6  &     4.03   &    36.5  & 19.9    &     0.495 &   &  MS1407.9+5954  &  & 0.99\\
 1REXJ141029+2820.9  &     4.66   &    28.8  & 19.2    &     --    & Y &                 &  & 0.83\\
 1REXJ141756+2543.3  &   136.81   &    89.6  & 17.5    &     0.237 &   &  HB891415+259   &  & 0.04\\
 1REXJ142645+2415.3  &     4.05   &     6.4  & 19.6    &     0.055$^t$&Y&                 &  & 0.98\\
 1REXJ143546+5815.3  &     9.73   &    26.4  & 19.7    &     0.299 & Y &                 &  & 0.40\\
 1REXJ163124+4217.0  &    29.07   &     7.4  & 19.3    &     0.468 &   &  16313+4217     &  & 0.62\\
 1REXJ223301+1336.0  &    28.45   &    25.1  & 19.4    &     0.214 & Y & RX J2233.0+1335 &  & 0.29\\
 1REXJ234053+8015.2  &    16.66   &    48.9  & 18.9$^E$&     0.274 & Y &                 &  & 0.20\\
 1REXJ234538$-$1449.4&     4.72   &    10.7  & 18.5    &     0.224 &   &  1ES2343-151    &  & 0.82\\
 1REXJ235320$-$1458.9&    14.95   &    15.0  & 19.7    &     --    & Y &                 &  & 0.42\\
\hline
\end{tabular}\\
\end{center}
\footnotesize{}
Column~1: Name, based on the radio position;

Column~2: The unabsorbed X-ray flux in the 0.5-2.0~keV energy 
band; 

Column~3: The integrated radio flux density at 1.4~GHz 
(from the NVSS catalog);

Column~4: The optical blue magnitude. We have used preferentially the 
APM O magnitude. 

When not available, a blue magnitude has been estimated 
in the following ways:  

($^E$) using the APM E magnitude, as described in the text

($^V$) using the V magnitude published in Padovani \& Giommi (1995) 
assuming B--V = 0.6 (mean value for the BL Lacs). 

($^B$) using the B magnitude published in Nass et al. (1996)

Column~5: The redshift. Tentative redshifts are indicated with ($^t$).
The redshifts are either computed on the basis of our spectra or they
are taken from NED. 

($^1$) Bauer et al. (2000) reports z=0.2106

($^2$) Cao, Wei \& Hu (1999) reports z=0.27

($^3$) redshift taken from Padovani \& Giommi (1995). Different redshift
in NED (z=0.13)

($^4$) redshift taken from Padovani \& Giommi (1995). Different redshift 
in NED (z=0.182) taken from Bade et al. (1996);

($^5$) Bauer et al. (2000) reports z=0.469

Column~6: A flag indicating that the object has
been observed spectroscopically by us (flag=``Y''). 

The flag
``Y'' does not mean necessarily that the object is newly
discovered. If a classification as BL Lac is already present
in the literature an alternative name is reported in
column~7.

Column~7: An alternative name from the literature, when available.

Column~8: type of BL Lac: ``BLC''= BL Lac candidate, i.e. the
     value of $\Delta$ is between 25\% and 40\%.

\end{minipage}
\vspace{3mm}

\small



\begin{references}{}
 \reference{}Avni, Y.,  Bahcall, J.N. 1980, \apj, 235, 694
 \reference{}Bade, N., Beckmann, V., Douglas, N. G., Barthel, P. D., Engels, 
  D., Cordis, L., Nass, P., \& Voges, W. 1998, A\&A, 334, 459
 \reference{}Bauer, F. E., Condon, J.J., Thuan, T.X., Broderick, J.J.
	2000, ApJS, 129, 547
 \reference{}B\"ottcher, M., Dermer, C.D. 2001, ApJ, in press
 \reference{}Caccianiga, A., Maccacaro, T., Wolter, A.,
             Della Ceca, R., Gioia, I.M. 2001, in ASP Conf.Ser., 
                Blazar Demographics and Physics,
                ed. P. Padovani \& C.M. Urry, (San Francisco, ASP), 238
 \reference{}Caccianiga, A., Maccacaro, T., Wolter, A.,
             Della Ceca, R., Gioia, I.M. 2000, A\&AS, 144, 247
 \reference{}Caccianiga, A., Maccacaro, T., Wolter, A.,
             Della Ceca, R., Gioia, I.M. 1999, \apj, 513, 51 (Paper~I)
 \reference{}Caccianiga, A., March\~a, M.J.M., Ant\'on, A., Mack, K.-H., 
		Neeser, M. 2001, MNRAS, submitted
 \reference{}Cao, L., Wei, J.-Y., Hu, J.-Y. 1999, A\&AS, 135, 243
 \reference{}Cavaliere, A., Malquori, D. 1999, \apj, 516, 9
 \reference{}Cavaliere, A., D'Elia, V. 2001, \apj, submitted
 \reference{}Condon, J.J.,  Cotton, W.D., Greisen, E.W., 
            Yin, Q.F., Perley, R.A., Taylor, G.B., Broderick, J.J. 1998,
            \aj, 115, 1693
 \reference{}Della Ceca, R., Zamorani, G., Maccacaro, T., Wolter, A.,
             Griffiths, R., Stocke, J.T., Setti, G. 1994, ApJ, 430, 533
  \reference{}Falomo, R., Scarpa, R., Bersanelli, M. 1994, ApJS, 93, 125
  \reference{}Gioia, I.M., Maccacaro, T., Schild, R.E., 
        Wolter, A., Stocke, J.T., Morris, S.L., Henry, J.P. 1990, 
                ApJS, 72, 567
  \reference{}Giommi, P., Pellizzoni, A., Perri, M., Padovani, P. 2001,
	in ASP Conf.Ser., Blazar Demographics and Physics,
        ed. P. Padovani \& C.M. Urry, (San Francisco, ASP), 227
  \reference{}Giommi, P., Menna, M.T., Padovani, P. 1999, \mnras, 310, 465
  \reference{}Laurent-Muehleisen, S. A., Kollgaard, R. I., Ciardullo, R. 
  Feigelson, E. D., Brinkmann, W., \& Siebert, J. 1998, \apjs, 118, 127
 \reference{}Maccacaro, T., Caccianiga, A., Della Ceca, A., 
             Wolter, A., Gioia, I.M. 1998, AN, 319, 15
  \reference{}Maccacaro, T., et al. 1982, \apj, 253, 504
  \reference{}Maccacaro, T., Gioia, I.M., Maccagni, D., Stocke, J.T.
                1984, \apj, 284, L23
  \reference{}March\~a, M. J. M., Caccianiga, A., Browne, I. W. A., 
		Jackson, N. 2001, MNRAS, 326, 1455
  \reference{}March\~a, M. J. M., Browne, I. W. A., Impey, C. D., \& Smith, 
  P. S. 1996, \mnras, 281, 425
  \reference{}Morris, S. M., Stocke, J. T., Gioia, I. M., Schild, E., 
       Wolter, A., Maccacaro, T., \& Della Ceca, R. 1991, \apj, 380, 49
  \reference{}Nass, P., Bade, N., Kollgaard, R. I., Laurent-Muehleisen, S. A.,
              Reimers, D., Voges, W. 1996, 309, 419
  \reference{}Padovani, P. 1993, MNRAS, 263, 461
  \reference{}Padovani, P. 2001, in ASP Conf.Ser., 
                Blazar Demographics and Physics,
                ed. P. Padovani \& C.M. Urry, (San Francisco, ASP), 163
  \reference{}Padovani, P., \& Giommi, P. 1995, MNRAS, 277, 1477
  \reference{}Paturel, G., Garnier, R., Petit, C., Marthinet, M.C. 1996, 
                A\&A, 311, 12
  \reference{}Perlman, E., Padovani, P., Giommi, P., Sambruna, R., 
             Jones, L. R., Tzioumis, A., \& Reynolds, J. 1998, \aj, 115, 1253
  \reference{}Rector, T. A.\& Stocke, J.T. 2001, \aj, 122, 565
  \reference{}Rector, T. A., Stocke, J.T., Perlman, E.S., Morris, S.L., 
             Gioia, I.M. 2000, \aj, 120, 1626
  \reference{}Stickel, M., Padovani, P., Urry, C. M., Fried, J. W., 
      \& K\"uhr, H. 1991, \apj, 374, 431
  \reference{}Stocke, J.T., Morris, S.L., Gioia, I.M., Maccacaro, T., 
                Schild, R.E., Wolter, A. 1989, in BL Lac objects, Vol. 334, 
		ed. L. Maraschi, T. Maccacaro, M.-H. Ulrich (Berlin: 
		Springer-Verlag), 242
  \reference{}Stocke, J.T., Morris, S. L., Gioia, I.M.;
               Maccacaro, T., Schild, R., Wolter, A., Fleming, T.A., 
               Henry, J.P. 1991, ApJS, 76, 813
  \reference{}Wolter, A., Ciliegi, P., della Ceca, R., Gioia, I. M.,
               Giommi, P., Henry, J. P., Maccacaro, T., Padovani, P.,
               Ruscica, C. 1997, MNRAS, 284, 225
  \reference{}Wolter, A., Ruscica, C., \& Caccianiga, A. 1998, \mnras, 299, 
                1047
  \reference{}Wolter, A., Caccianiga, A., Della Ceca, R., \& Maccacaro, T. 
                1994, \apj, 433, 29
  \reference{}Wolter, A., Gioia, I.M., Maccacaro, T., Morris, S.L., 
                Stocke, J.T. 1991, ApJ, 369, 314

\end{references}
\end{document}